\documentclass[final, runningheads]{llncs}
\usepackage[T1]{fontenc}
\usepackage{graphicx}
\usepackage{amsmath}
\newcommand{\itadata}{\footnotesize \textsl{Workshop Scientific HPC in the pre-Exascale era (part of ITADATA2024)}}
\usepackage{fancyhdr}
\fancypagestyle{empty}{\fancyhf{}\fancyhead[C]{\itadata}
 \fancyhead[R]{\includegraphics[width=.05\textwidth]{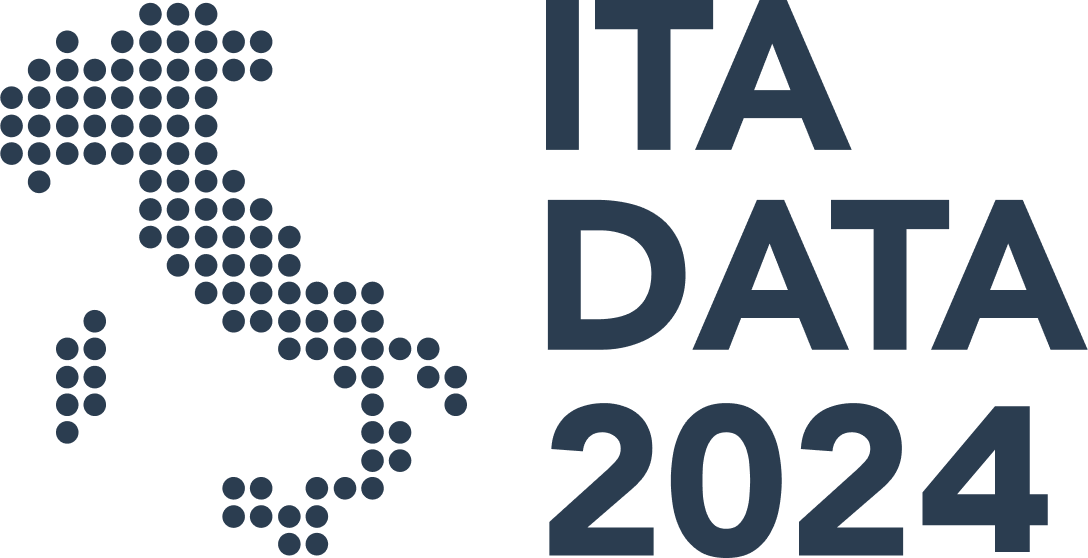}}}

\makeatletter
\begin{document}
\title{Computational Challenges of 21st Century Global Astrometry}
\author{Beatrice Bucciarelli\orcidID{0000-0002-5303-0268}}
\authorrunning{F. Author et al.}
\institute{ INAF Astrophysical Observatory of Turin, Italy\\
\email{beatrice.bucciarelli@inaf.it}}
\maketitle              \begin{abstract}
Major advancements in space science and detector technology brought
about a revolution in global astrometry, the science of measuring
distances and motions of stars in the Milky Way and in the local
universe. From the first ESA astrometric mission HIPPARCOS of the
early 80s to the current Gaia mission, the data volume and
computational complexity of the full reduction process has increased
by several orders of magnitude, requiring high-performance computing
and data throughput. We review the principles and computational
complexity of general global astrometric models that lead to the statistical
treatment of an extra-large, highly non-linear estimation
problem. Some numerical aspects of inspecting Gaia's proper motions to
find cosmological signals at all scales are also addressed.
\end{abstract}
\section{Astrometry at a glance}
The history of astrometry is rooted in the ancient past, when the
movements of stars and planets were recorded through naked eye
observations. Thanks to telescope and detectors improvements, star
positions have been measured with progressively increasing accuracy,
which, however, inevitably suffered from the perturbative effect of the Earth's
atmosphere.

The impasse was overcome once astrometric observations
started to be conducted from space in the early '90s. Since
then, the drastic gain in measurement accuracies, supported
by progress in  theoretical modeling and data reduction capabilities,
has put the scientific impact of the new astrometry at the forefront of astrophiscs.

\subsection{Objectives}
The study of motions of celestial objects through space and
time is the main concern of astrometry, with direct
implications on the knowledge of our Universe, from the outskirts
of the Sun all the way to cosmological scales.

A very tiny motion of stars which deserves particular attention is their
yearly apparent displacement due to the motion of
the Earth around the Sun. The angular semi-amplitude
of such displacement, called stellar {\it parallax},  is the angle under
which the radius of the Earth orbit is seen from the star.
The parallax is smaller than 1 arcsecond (1 arcsecond = 1/3600 of
degree) for all stars, and it is
inversely proportional to the star distance
from the Sun.
Knowing the distance to a star is of fundamental importance to astrophysics as it
represents the only way to transform apparent measurements into actual
physical quantities. Astrometrically derived parallaxes are also the
basic calibrators of the cosmic distance scale.
The stars' intrinsic motion, called {\it proper motion}, is another
fundamental astrometric parameter which, complemented by the
spectroscopic determination of radial velocityies, is the basic
material to study the kinematics and dynamics of stellar systems and investigate
the formation and evolution of the Galaxy. Furthermore, obtaining highly accurate
astrometry of extragalactic objects, which can be achieved from space, is instrumental to the realization of a {\it
  non-rotating} celestial reference frame. The presence of residual
proper motions of extragalactic objects whose peculiar motions are too
small to be detected, can be exploited for the search of astrometric signatures of cosmological origin.

\subsection{Methods}

The basic measurement of astrometry consists in the angle between the
direction of incoming light from source pairs in the sky.  Observations
can be made from ground and from space, depending upon the
instrumental techniques and the typical field of view (FOV). In general,
one can distinguish three classes of astrometry: small-field
(fraction-of-degree FOV; double or multiple stars, clusters);
large-field (few-degree FOV; position of celestial bodies with respect
to reference stars); global-astrometry (stars all over the sky producing a
consistent set of positions on the celestial sphere), only obtainable
from space because no observatory on the ground can see the entire sky
\cite{12}.

\subsubsection{Relative vs absolute parallaxes}
Traditional small-field astrometry only measures
{\it relative} parallaxes. This is because the parallax factor ($\approx
\sin\theta$, where $\theta$ is
the angle between the star and the Sun
as seen by the observer) is virtually the same for all the stars in
the field so that the single parallax is uniquely determined up to
an additive constant. Hence, the relative parallax requires an additive correction that
must be determined by other methods. From space, {\it absolute} parallaxes,
thereby distances, can be obtained by observing stars
that are widely separated on the celestial sphere, and therefore have
very different parallax factors.

\section{Global space astrometry}
The first space experiment dedicated to astrometry was the ESA's HIPPARCOS
mission \cite{7}, which introduced a radically different technique,
i.e. scanning space astrometry. One of the aims of {\it scanning} space
astrometry is to build a globally consistent celestial reference frame.
The realization of the reference frame consists of a set of objects, a
theory (a model) governing  their motion in space, and
a set of coordinates (a {\it catalog)} specifying the motion of each
individual object through its 5 astrometric parameters (positions,
proper motions, and parallax).
\subsection{Basic principles}
A continuously scanning telescope transforms positional information
into {\it timing} data. That is, it determines the precise time when the
centre of a star image has some well-defined position in the FOV. The
resulting ‘observation time’ is a one-dimensional (along-scan, AL)
measurement of stellar position relative to the instrument axes.

Ideally, the angle between two objects could be obtained by
orienting the instrument such that both objects are precisely on the
instantaneous scanning great circle. However, what is measured is the
projection of the desired angle onto the scanning circle, and given
the size of the FOV, 
the tolerance for AC errors may be up to 100 times more relaxed than
AL \cite{10}.
So, the design of a scanning telescope is optimized for measurements in
the AL direction, getting at the same time an approximate position of the star in the
across-scan (AC) direction but with a much less precision. 
The final astrometric catalogue is built up from a very large number of such
observation times, by a process that involves also a precise
reconstruction of the instrument
pointing (attitude) as a function of time and of the optical mapping of the CCDs through the telescope into the celestial sphere.

\subsubsection{The HIPPARCOS mission}
The HIPPARCOS satellite was launched in August 1989 and remained in operation
until 1993; the mission primary goal was to measure  the 5 astrometric
parameters of $\approx$ 120000 program stars up
to a magnitude of $\approx$ 12 with a precision of $\approx$ 2-4 milli-arcseconds.
The satellite, by means of a double mirror (the beam combiner)
simultaneously observed two  small patches  (0.9°x0.9°, the FOV) about
58° (the so-called {\it basic angle}) apart in the sky.
The data reduction process was at the time a formidable adjustment
problem, which was undertaken by the FAST and NDAC Consortia.
A direct solution approach by elimination of stellar unknowns, 
would have required about $n^3/3 \approx 10^{17}$ flops and the
administration of $n^2/2 \approx 5\cdot 10^{11}$ double-precision
reals (where $n\approx 10^6$ calibration parameters), a non-trivial task even for
supercomputers and parallel processing resources in the
mid-eighties. 
To meet the computer resources available, a three-step method was
devised \cite{9}. Measurements taken over an interval of about 10 hours,
corresponding to the orbital period of the satellite in its revised
elliptical orbit, were combined together. The first part of the
adopted 'three-step' analysis method consisted of estimating the
relative one-dimensional positions of all the stars contained in the
strip of sky scanned over the 10-hour interval (roughly a circle of 1
degree width), each with respect to the others.
The second step, called the sphere solution in Hipparcos terminology, was to use the abscissae determined from the great-circle reduction  for the calculation of the 5 astrometric parameters of the primary stars. The final part of the three-step method consisted in the estimation of the astrometric parameters of all the remaining non-primary stars.
The so-called sphere solution involved $\approx 400000$ stellar unknowns (5 times the number of primary
stars) and ~2M calibration parameters (the number of attitude coefficients and instrument parameters).
Two basic algorithms were developed within the FAST Consortium \cite{5}
to perform the sphere solution: one was to solve the normal system taking into account its block structure (2x2 Block Cholesky factorization or 2-Block SOR iterative method); the other was to solve the condition equation system by means of an iterative gradient-type method.
The LSQR procedure, an iterative algorithm based on the Lanczos method
was first successfully used for the sphere solution. This was achieved
by using specific vectorial optimization of the initial code, taking into account the
vectorial features of the Cray X-MP/12 at the supercomputing center
CINECA in Bologna, I.

\begin{figure}
\begin{center}
\includegraphics[width=0.7\textwidth]{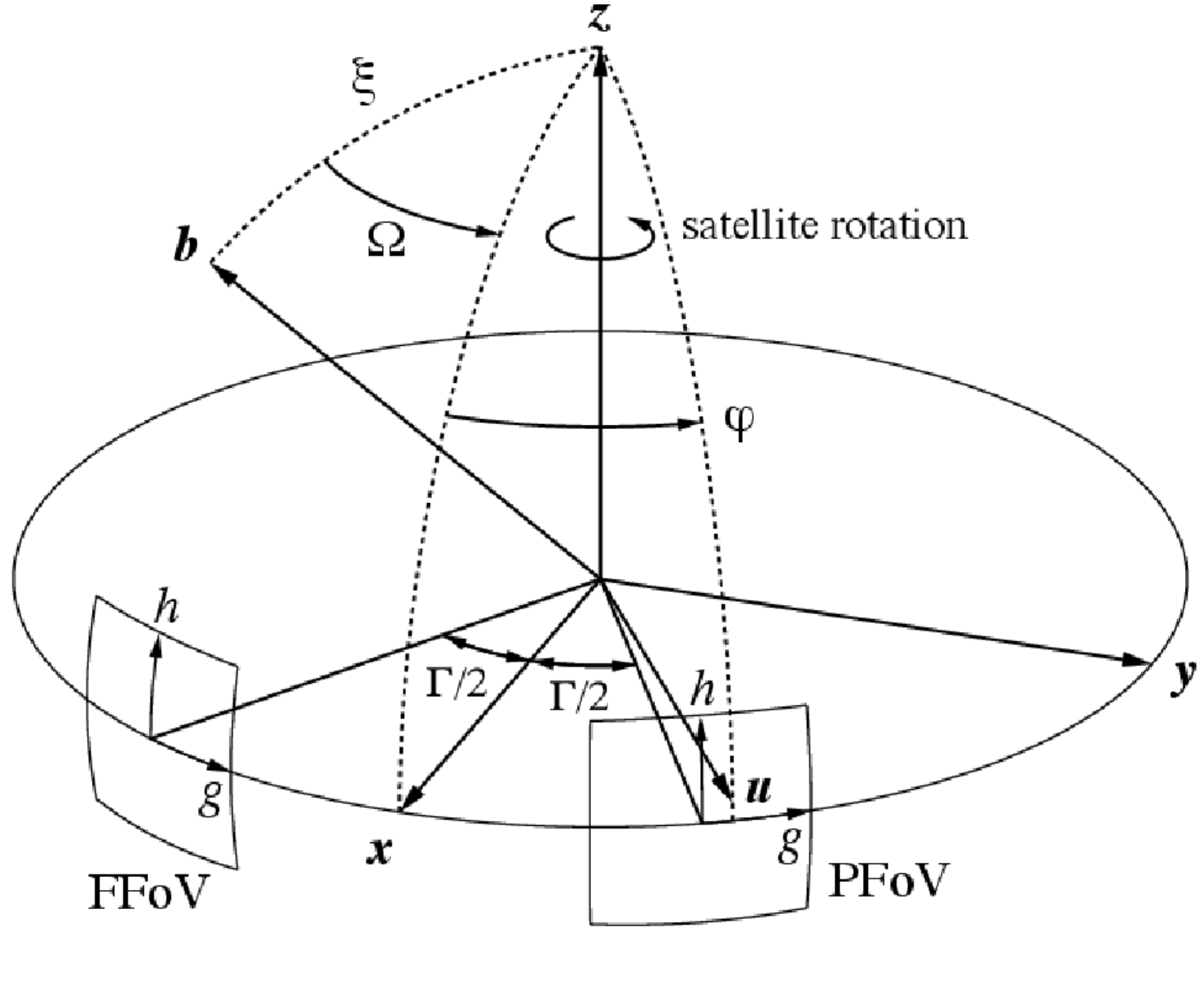}
\caption{Gaia satellite scanning law and observation mode: axes ({\bf x},{\bf y},{\bf z}) define the Satellite Reference
 System (SRS); the {\bf x} axis rotates around the spin axis
 {\bf z} with a 6h period rotation angle;
 coordinates ({\it g},{\it h}) are the along (AL) and across (AC) scan direction
 on the istantaneous scanning great circle, $\Gamma$ is the Basic Angle separating
 the preceding (PFOV) and following FFOV) field of views. To obtain
 full-sky coverage, the spin axis {\bf z} precesses with a period
 of 63 days around the Earth-Sun direction with a fixed aspect angle
 $\xi$, (from Butkevitch et al. \cite{2})} \label{fig1}
\end{center}
\end{figure}

\subsubsection{The Gaia mission}
The Gaia satellite was launched on December 13 2013 from the French Guiana, in a
Lissajous-type orbit around the lagrangian point L2 of the
Sun-earth/moon system (1,500,000 km from earth) corotating with the
Earth in the 1-year orbit around the Sun \cite{4} .The Gaia Data Processing and Analysis Consortium (DPAC) processes the raw data that are published in one of the largest stellar catalogues ever made.
Gaia is conducting an optical all-sky survey of the Milky Way down to magnitude 20:
near 2 billion objects, constituting about 1\% of all the
stars populating the Milly Way.  
Gaia operation and observing principles are based on those of its
predecessor mission HIPPARCOS, but its performance is  ~3 order of
magnitude more accurate.

The measurement principle relies on the
systematic, repeated observation of star positions in two fields of
view (FOV) separated by a $106.5^{\circ}$ angle, the {\it Basic
 Angle}, see figure \ref{fig1}. For this, the spacecraft is slowly rotating at an angular rate of 1° per minute around an axis perpendicular to those FOVs, describing a great circle in 6 hours. The scan axis further describes a slow precession motion around the sun-to-earth direction with a period of 63 days. Full-sky coverage is hence obtained after 3 months. 
The focal plane assembly is the largest ever developed for a space
application, with 106 CCD-s, a total of almost 1,000 million pixels
and a physical dimension of 1m x 0.4m. The detectors operate in a Time Delay
Integration mode (TDI) where the photoelectrons generated by the star
image are clocked across a CCD together with the moving star image.
Each CCD features 4500 pixels in the along-scan and 1966 in the
across-scan directions. The relation between the size of the PSF and that of the
CCD pixel is fundamental for achieving the measurement accuracy. The
best compromise was found with a 35m focal length and pixel sizes of
10 microns (59 milli-arcseconds) along the star-crossing direction and 30 microns
perpendicular to it.

\section{The Sphere Solution in the Gaia mission}
Gaia's observation is represented by the precisely estimated
instant when the star image centre crosses the CCD ‘observation line’
(nominally situated between the 2250th and the 2251st pixel). This
instant is called the CCD observation time. The timing observations
provide accurate (typically 0.1 to 1 milli-arcseconds) information about the
instantaneous relative along-scan position of the observed
objects.

There are about $10^8$ primary stars, each observed about 80 times in 5
years, i.e. $\approx$720 CCD transits. So, assuming one hundred
millions stars, there will be $7.2\cdot 10^{10}$ transit times (about 45 stars
are observed each second). Moreover, the attitude parameters to be
estimated over 5 years are about $4\cdot 10^7$.
The number of instrumental calibration parameters is $\approx 10^6$,
estimating large and medium-scale effects which model the physical
geometry of CCDs and optical aberrations. In its most general formulation, the Gaia astrometric solution can be
considered as the minimization problem
\begin{equation}
\text{min}_{(\bf{s},\bf{n})}||\bf{f}^{obs}-\bf{f}(\bf{s},\bf{n})||
\end{equation}
where {\bf s} is the vector of unknown stellar parameters describing
the barycentric motion of the stars and {\bf n} a vector of “nuisance
parameters” describing the instrument, not of direct interest for the
stellar problem but required for the modeling of the data.
$\bf{f}^{obs}$ represents the vector of observations and
{\bf f}({\bf s},{\bf n}) is the observation model, i.e,
the expected detector coordinates calculated as function of the
astrometric and nuisance parameters.
The minimization problem corresponds to the least-squares solution of
the overdetermined system of equations
\begin{equation}
f_l^{obs}=f_l(\vec{s}_i,\vec{n}_j), l=1,...\text{number\;of\;observations}
\end{equation}
where {\it i} indicates the stellar source and {\it j} the segment of
attitude parameters pertaining to the {\it i}th source.
The function {\bf f} is highly non-linear in {\bf s} and {\bf n}. Nonetheless, thanks to the data processing prior to the astrometric solution, the initial errors in these parameters are small, and second-order terms of the linearized equations are typically less than
$10^{-12}$ rad ($\approx$ 0.2 micro-arcseconds, which is
negligible in comparison with the noise of a single observation. So,
the equation is linearized around some suitable reference values:
\begin{equation}
f_l^{obs}-f_l^{calc}=\frac{\partial f_l}{\partial{\bf s}_i}{\bf x}_{si}+\frac{\partial f_l}{\partial{\bf n}_i}{\bf x}_{nj}
\end{equation} 
The weighted least-squares system is finally obtained multiplying each equation by the square root of its statistical weight. 

\subsection{The minimization problem}
Given the $\approx10^8$ primary sources of Gaia, the number of
unknowns in the global minimization process is $5\cdot10^8$ for the
sources, $4\cdot10^7$
and $\approx10^6$ for the calibration parameters.
The number of elementary observations to be considered is about $8\cdot10^{10}$.
The size of the dataset and the large number of parameters would not
by themselves be a problem if the observations could be processed
sequentially; however, as it will be exemplified in the following, this is not the case.

\subsubsection{Matrix structure}
Using matrix notation, the linearized observation equation can be
written as $O{\bf x}={\bf b}$. Considering for simplicity only astrometric and
attitude unknown ${\bf x}={\bf x}({\bf s},{\bf a})$, and sorting all the
observations by source one gets what is called a non-square block
angular matrix of the form

\begin{gather}
 \begin{bmatrix} S_{1} & 0 & \ldots & 0 & A_{1} \\
   0 & S_{2} & \ldots & 0 & A_{2} \\
   \vdots & \vdots & \ddots & \vdots & \vdots \\
   0 & 0 & \ldots & S_{n}  & A_{n}
 \end{bmatrix}
 \begin{pmatrix}
 {\bf x}_{s1} \\ \vdots \\ {\bf x}_{sn} \\ {\bf x}_a
\end{pmatrix}
=
\begin{pmatrix}
{\bf  b}_{s1} \\ \vdots \\ {\bf b}_{sn}
\end{pmatrix}
\end{gather}
with {\it n} the number of primary sources. Being {\it m} the number
of attitude unknowns and $o_{si}$ the number of observations of the
i-th source, the dimensions of the different sub-matrices and sub-vectors  are $\text{dim}(S_i)= (o_{si}\times 5)$, $\text{dim(}A_i)=
(o_{si}\times m)$, $\text{dim}(x_{si})= (5\times 1)$, $\text{dim}(x_a)= (m\times
1)$, $\text{dim}( b_{si})= (o_{si} \times 1)$ .
$S_i$ are full matrices, while $A_i$ are very sparse with
$[A_i]_{\alpha\beta}\neq 0$ only if the $\alpha$th observation of source {\it i} is
linked to the $\beta$th attitude parameter.

\subsubsection{Computational complexity}
The difficulty of handling such a large number of
parameters is caused by the strong connectivity among the
observations which prevent a sequential treatment of the data.
In fact, each source is observed relative
to a large number of other sources simultaneously in the 2 FOVs,
linked together by the attitude (and calibration) model. 

The sparseness structure of the matrix is directly related to the
choice of functions representing the attitude. 
The Gaia attitude is modelled as a spline, i.e.,  a piecewise
polynomial function, defined on some time
interval, that can be written as the linear combination of basis
functions called B-splines. The latter  are defined in by a non-decreasing
sequence of M+1 time knots (M=4 for cubic B-splines), so that for any time
$t_i$ there are 4 non-zero cubic B-splines, and the associated spline
coefficients will be   $a_{(i-M+1)},a_{(i-M+2)},…,a_i$. Therefore, the
sub-vector $a_j$ of the attitude parameters will consist of 3M
scalar values, namely M spline coefficients for each of
the three orientation angles of the satellite axes.
So, the observation equation for different sources involves disjoint
source parameters sub-vectors $\vec{s}_i$ but may refer to the same attitude
sub-vector $\vec{a}_j$. Then, the fraction of non-zero elements of A is equal to 3M/m (with M equal to
the B-spline order), and for the full condition matrix $O$ the fill
factor (i.e. the number of non-zero elements) is (5+3M)/(5n+m), which for $n=10^8$, $m=4\cdot10^7$, and $M=4$ is
$\approx 2\cdot 10^{-8}$.

\subsubsection{Numerical approach}
The least-squares problem is classically solved by forming the normal equations $
O^T O{\bf x}=O^T {\bf b}$, $( O^T O\equiv  N)$;  then if the normal matrix
$N$ is invertible, the solution is simply
${\bf x}=(O^TO)^{-1} O^T {\bf b}$. The structure of the normal matrix
is then
\begin{gather}
 \begin{bmatrix} S_1^TS_1 & 0 & \ldots & 0 & S_1^TA_1 \\
   0 & S_2^TS_2 & \ldots & 0 & S^T_2A_2 \\
   \vdots & \vdots & \ddots & \vdots & \vdots \\
   0 & 0 & \ldots & S^T_nS_{n}  & S^T_nA_{n}\\
   A^T_1S_1 & A^T_2S_2 & \ldots & A^T_nS_n & \sum A^T_iA_i
 \end{bmatrix}
 \begin{pmatrix}
 {\bf x}_{s1} \\ {\bf x}_{s2} \\\vdots \\ {\bf x}_{sn} \\ {\bf x}_a
\end{pmatrix}
=
\begin{pmatrix}
 S^T_1{\bf b}_{s1} \\ S^T_2{\bf b}_{s2} \\ \vdots \\ S^T_n{\bf
   b}_{sn} \\ \sum A^T_i{\bf b}_{si}
\end{pmatrix}
\end{gather}
The full normal equation is symmetric of size $5n+m$, and has a doubly
bordered block diagonal form, with a block size of 5 and border width
{\it m}. 
The dimensions of the sub-matrices are:  $S_i^T S_i (5\time 5)$,
$S_i^T A_i (5\times m)$,
$\sum A_i^T A_i (m\times m)$, the fill
factor of $\sum A_i^T A_i (m\times m)$ is $\approx 3(2M-1)/m$,
i.e., $\approx 10^{-6}$, and that of $N$ is $\approx 310/n
\approx 3\cdot 10^{-6}$. 

A standard way to handle the normal equations with the
block-diagonal-bordered structure is to successively eliminate the
unknowns along the block diagonal, leaving a reduced normal equation
system for the remaning unknowns, i.e., the attitude parameters. The gain
is a huge reduction in the size of the system, at the expense of a
much denser {\it reduced} normal matrix. Bombrun et al.
\cite{1} used numerical
simulations to study the complexity of the Cholesky factorization of
the reduced normal matrix: 
\[
R_a=\sum_{i=1}^n(A_i^T A_i-A_i^T S_i (S_i^T S_i )^{-1} S_i^T A_i ) 
\]
and showed that a direct solution of the reduced normal equations for
a 5 year mission (7300 spin periods) would require about $1.3 \cdot
10^{21}$ flops, the operation count for the
Cholesky decomposition of a $m\times m$ matrix, i.e.,
$\approx m^3/6$ (having ignored calibration and global unknowns). Also, an upper triangular matrix with the dimension of the reduced normal matrix will require around 2 million GigaBytes. So, with current limitations in terms of storage and floating-point operations, a direct method cannot be practically used to solve the Gaia astrometric adjustment problem.

Gaia's pipeline solution uses a block-iterative technique, where a
rigorous solution of the normal system is obtained by solving
separately each block of unknowns and iterating the process until
convergence \cite{11}. 
However, the associated uncertainties are only approximately estimate;
e.g., for a given source, the formal standard errors of the 5
astrometric parameters are obtained from the diagonal elements of the
inverse of the corresponding 5x5 blocks of the normal matrix, which neglects the statistical correlations introduced by the attitude and calibration models, as discussed in Holl and Lindegren \cite{8}. This causes the actual uncertainties to be underestimated.
A different solution method implemented by the Astrometric Verification Unit
(AVU, Vecchiato et al. \cite{15}.) makes use of LSQR \cite{14},
a conjugate-gradient-like iterative algorithm, especially suited to handle large and sparse linear
system, and already proved effective for the HIPPARCOS sphere
reconstruction. Given the differential nature of the observations, the sphere solution
system has a rank defect of 6. The 6-dimensional null space is in fact
well known and corresponds to the undefined orientation and spin of the
reference system in which both the source parameters and the attitude
are expressed. While the block-iterative technique naturally brakes the
degeneracy of the problem, the LSQR method, if no constraints are enforced,
converges to the minimum norm solution. It also provides an
estimation of the standard deviations of all the unknowns, and we have
modified it from its original version in order to perform the covariance estimation of any selected group of unknowns (the
off-diagonal elements of the inverse normal matrix). Given the
data volume and computational complexity of the problem,
the employment of High Performance Computing (HPC) techniques was
essential \cite{3}. 

\section{Cosmological signatures}

\subsection{Extragalactic proper motions}
Extragalactic proper motions can reveal a variety of cosmological and
observer-induced phenomena over a range of angular scales
(Darling et al.  2019), e.g., Cosmic aberration drift, gravitational waves, anisotropic expansion.
\subsection{Vector spherical harmonic analysis}
Global and local features of  the residual proper motion vector field
$\vec{\mu}$ of quasars (QSOs) can be analysed by decomposition into
a set of orthogonal basis functions called Vector Spherical
Harmonics (VSH), i.e.
\begin{equation}
\vec{\mu}=\sum_{l,m}(t_{lm}\vec{T}_{lm}+s_{lm}\vec{S}_{lm})
\end{equation}
where $\vec{T}_{lm}$ and $\vec{S}_{lm}$ are respectively the toroidal (or {\it magnetic})
and spheroidal (or {\it electric}) base functions of degree {\it l}
and order {\it m}; {\it t} and {\it s} are the coefficients of the
expansion to be estimated. Projection of  {\bf T} and {\bf S} onto
the normal unit vectors $\vec{e}_\alpha$, $\vec{e}_\delta$ of the
local frame associated to the spherical coordinates $(\alpha,\delta)$ gives:
\begin{equation}
  \begin{aligned}
 \vec{T}_{lm}=\frac{1}{\sqrt{l(l+1)}}\left[\frac{\partial Y_{lm}}{\partial\delta}
 \vec{e}_\alpha-\frac{1}{\cos\delta}\frac{\partial
   Y_{lm}}{\partial\alpha}\vec{e}_\delta\right] \\
\vec{S}_{lm}=\frac{1}{\sqrt{l(l+1)}}\left[\frac{1}{\cos\delta}\frac{\partial Y_{lm}}{\partial\alpha}
 \vec{e}_\alpha+\frac{\partial
   Y_{lm}}{\partial\delta}\vec{e}_\delta\right]
\end{aligned}
\end{equation}
and $Y_{l,m} (\alpha,\delta)$ are the standard spherical harmonics.
\subsubsection{Numerical aspects}
The residual proper motion field is modeled as a vector field defined
on the surface of a sphere orthogonally to the radial direction, i.e.,
$\vec{V}(\alpha,\delta)$, where $(\alpha, \delta)$ are equatorial
coordinates. Such field can be expanded in a unique linear combination of VSH functions as
\begin{equation}
\vec{V}(\alpha,\delta)=\sum_{l=1}^{\infty}\sum_{m=-l}^{l}(t_{lm}\vec{T}_{lm}+s_{lm}\vec{S}_{lm})
\end{equation} 
where, practically, the expansion is truncated to a certain degree
{\it l} and the coefficients $t_{lm}$, $s_{lm}$ are estimated in a least-squares adjustment.  
Similarly to Fourier decomposition, as the degree {\it l} increases, smaller
details in the field systematics are reached, with angular resolution
$\theta \approx \pi/l$. Given $N$ sources and a truncation of the VSH decomposition to
a degree $L$, the dimensions of the system of condition equations are
$2N\times 2L(L+2)$.
The data storage of the full design matrix could be a problem, so it
is good practice to build up the normal matrix on the fly, taking also
advantage of its symmetry. For a large dataset, ie., $(N>>L^2)$ this
step is the most demanding in terms of computation time \cite{13}.

Signals due to residual rotation and acceleration (or {\it glide}) of the
reference frame materialized by QSOs
are fully contained in the first degree toroidal and spheroidal
VSH components respectively. In particular, the glide signal is an observer-induced
aberration effect, due to the non-linear velocity of the Sun
around the Galactic center, with magnitude $\approx 5$
micro-arcseconds, corresponding to a linear acceleration of the Sun 
of $\approx 10^{-10} m/s^2$.
Astrometric signatures can also be investigated in the context of
cosmology. In fact, a stochastic gravitational wave (GW) background
causes the apparent position of
QSOs to fluctuate with
angular deflection of the order of the GW amplitude, mimicking a
proper motion field $\vec{\mu}(\alpha, \delta)$. Gwinn \cite{6} has
noticed that the resulting squared proper motion field averaged over the sky $⟨\mu^2 ⟩$, can be directly related to
the energy density of the cosmological GW background and is mostly
captured in the development of quadrupolar (degree 2) VSH
functions. The amplitude of such signals is expected to be at the sub-micro-arcsecond
level, a difficult challenge for Gaia, but certainly within reach of
the next-generation astrometry missions.

\end{document}